\documentclass[3p,times,procedia]{elsarticle}
\usepackage{nupha_ecrc}

\volume{00}

\firstpage{1}

\journalname{Nuclear Physics A}

\runauth{Frithjof Karsch}

\jid{nupha}

\jnltitlelogo{Nuclear Physics A}

\usepackage{amssymb}
\usepackage[figuresright]{rotating}

\def\lsim{\raise0.3ex\hbox{$<$\kern-0.75em\raise-1.1ex\hbox{$\sim$}}}
\def\gsim{\raise0.3ex\hbox{$>$\kern-0.75em\raise-1.1ex\hbox{$\sim$}}}

\begin{document}

\begin{frontmatter}

\title{Conserved charge fluctuations at vanishing and non-vanishing chemical potential}
\author{Frithjof Karsch}
\address{Fakult\"at f\"ur Physik, Universit\"at Bielefeld, D-33615 Bielefeld,
Germany\\
and\\
Physics Department, Brookhaven National Laboratory, Upton, NY 11973, USA}

\begin{abstract}
Up to $6^{th}$ order cumulants of fluctuations of net baryon-number, 
net electric charge and net strangeness as well as correlations among 
these conserved charge fluctuations are now being calculated in lattice 
QCD. These cumulants provide a wealth of information on the properties 
of strong-interaction matter in the transition region from the low 
temperature hadronic phase to the quark-gluon plasma phase. They can
be used to quantify deviations from hadron resonance gas (HRG) model 
calculations which frequently are used to determine thermal conditions 
realized in heavy ion collision experiments.
Already some second order cumulants like the correlations between net 
baryon-number and net strangeness or net electric charge differ significantly 
at temperatures above 155 MeV in QCD and HRG model calculations. 
We show that these differences increase at
non-zero baryon chemical potential constraining the applicability range
of HRG model calculations to even smaller values of the temperature.

\end{abstract}


\end{frontmatter}

\section{Introduction}
\label{intro}
The central goal of the beam energy scan (BES) program at RHIC is to seek
evidence for the existence of a critical point in the phase
diagram of strong-interaction matter. The hope is to detect this
postulated second order 
phase transition point through the analysis of higher order cumulants
of net charge fluctuations. Maxima of cumulants of net charge fluctuations,
e.g. the $2^{nd}$ and $4^{th}$ order cumulants, trace the chiral 
crossover transition line at small values of the baryon chemical potential
and 
diverge at a critical point. 

In heavy ion experiments the observed
net charge fluctuations are expected to reflect thermal conditions
at the time of chemical-freeze out of various hadron species. 
If this freeze-out happens close to the pseudo-critical
line for the chiral transition of QCD, where thermal fluctuations are
large, the measured fluctuations
have a chance to be indicative for the divergent fluctuations that will 
appear at a critical point in the QCD phase diagram.

A crucial anchor point for this scenario is to establish the relation
between freeze-out and the QCD chiral transition at small or even
vanishing net baryon chemical potential. In this case reliable theoretical
calculations, based on Taylor series expansions in lattice QCD, exist
and can be confronted with experimental findings at the LHC as well as
the highest beam energies at RHIC. In experiments at the LHC one
can analyze moments of charge fluctuations at almost vanishing 
baryon chemical potential ($\mu_B$) which allows a direct comparison with
lattice QCD calculations performed at $\mu_B=0$.

\section{Thermal conditions at vanishing net baryon number}
\begin{figure}[t]
\begin{center}
\hspace*{-0.1cm}\includegraphics[width=84mm]{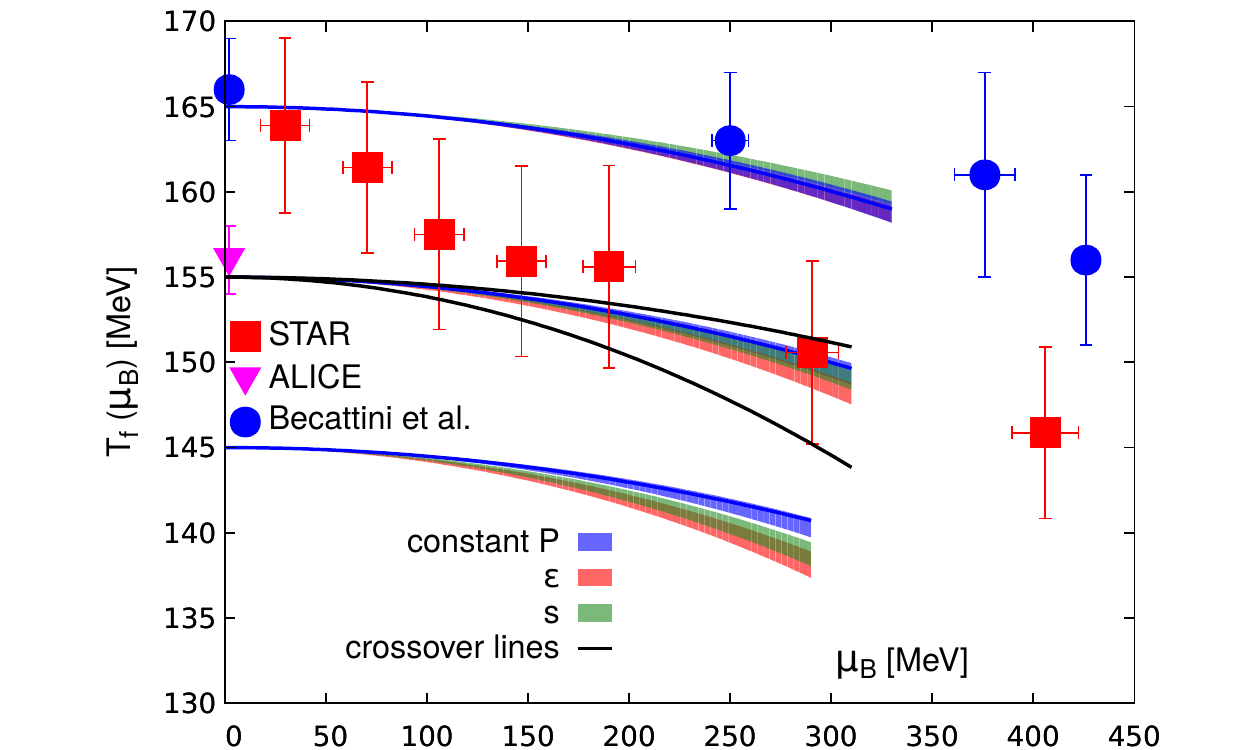}\hspace*{-0.4cm}
\includegraphics[width=72mm]{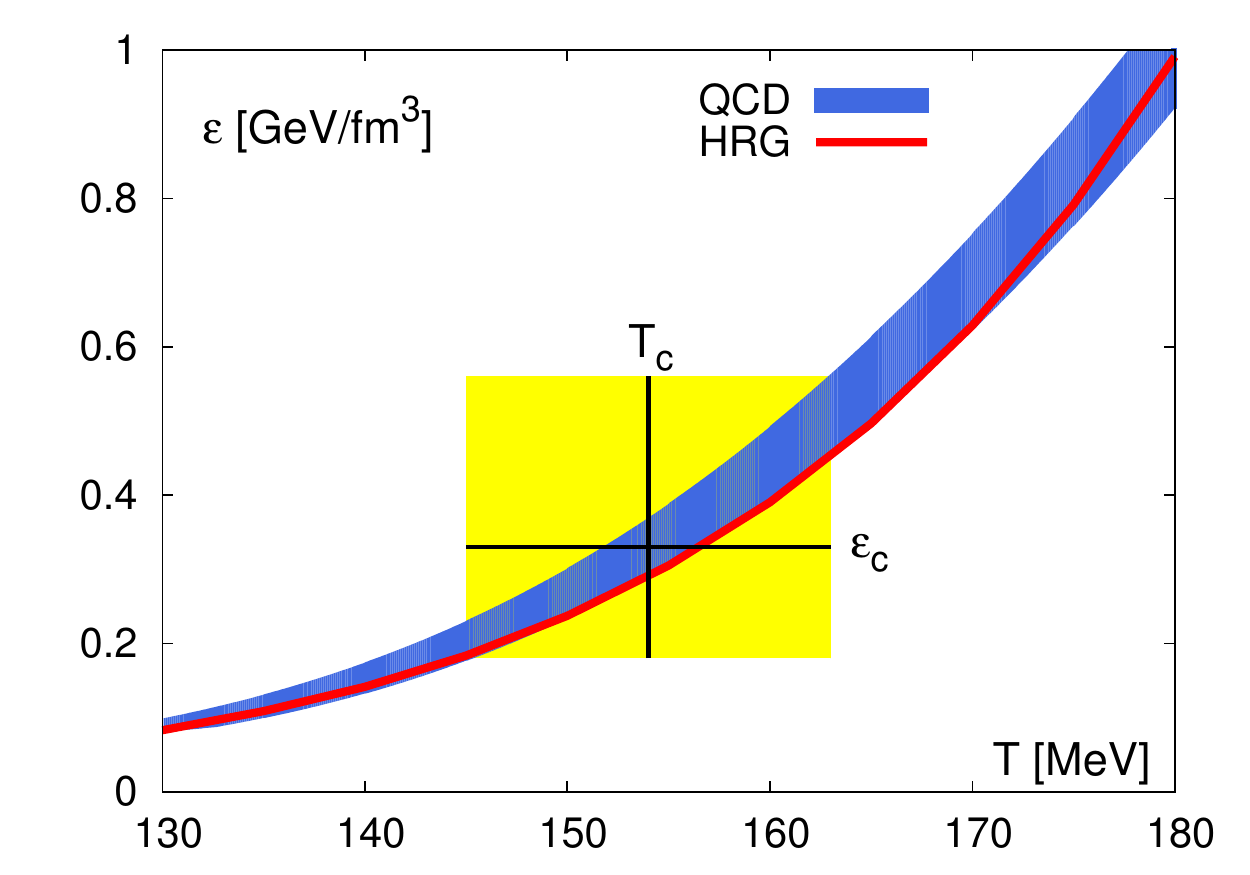}
\caption{{\it Left:} Lines of constant pressure, energy and entropy density,
as given in Table II of Ref.~\cite{Bazavov:2017dus},
as function of baryon chemical potential. Solid black lines indicate the
current uncertainty on the variation of the pseudo-critical temperature of 
the chiral transition, $T_c(\mu_B)$
with $\mu_B$. For a discussion of the data points see text.
{\it Right:}
The energy density at $\mu_B=0$ as function of temperature. The box
reflects current errors on the crossover transition temperature,
$T_c=154(9)$~MeV.
\vspace*{-0.5cm}
}
\label{fig:phasediagram}
\end{center}
\end{figure}

The basic bulk thermodynamic observables,
pressure ($P$), energy ($\epsilon$) and entropy ($s$) density
of strong-interaction matter
at vanishing baryon chemical potential, have been calculated in lattice
QCD (for a recent review see: \cite{Ding:2015ona}). These calculations have 
recently been extended to non-vanishing
baryon number densities using analytic continuation of calculations
performed at imaginary values of $\mu_B$ \cite{Gunther:2016vcp}
and Taylor expansions in $\mu_B$ \cite{Bazavov:2017dus}.
In Fig.~\ref{fig:phasediagram}~(left) we show results for lines of 
constant $P$, $\epsilon$ and $s$, in the $T$-$\mu_B$ plane
(phase diagram) obtained from a Taylor series up to 
${\cal O}(\mu_B^4)$ \cite{Bazavov:2017dus}. Lines are drawn for three values 
of these observables in the crossover region for the QCD chiral transition, 
which is well characterized by the current uncertainty on the chiral 
transition temperature, $T_c=154(9)$~MeV, at $\mu_B=0$ \cite{Bazavov:2011nk}.
As can be seen in Fig.~\ref{fig:phasediagram}~(right) in this temperature
interval the energy density
changes by about a factor three,
$\epsilon_c = (0.34\pm 0.16)$~GeV/fm$^3$ \cite{Bazavov:2014pvz}.

Also shown in Fig.~\ref{fig:phasediagram}~(left) are experimental 
results for freeze-out parameters determined by the ALICE Collaboration
at the LHC \cite{Floris:2014pta} and the STAR Collaboration from the BES
at RHIC \cite{Das:2014qca} by comparing measured particle
yields with predictions from a statistical hadronization model, which utilizes
the thermodynamics of a hadron resonance gas. Obviously, there is a 
significant difference in the determination of the freeze-out 
temperature at $\mu_B\simeq 0$.
While the ALICE result for the freeze-out temperature ($T_f$) agrees well with 
the central value of the pseudo-critical temperature ($T_c$), the STAR results
favor a larger value, $T_f\sim 165$~MeV, which is close to the 
hadronization temperature obtained by Becattini et al.
\cite{Becattini:2016xct}.

A 10~MeV accuracy for the determination of the freeze-out 
temperature, which anyhow is not considered to be a 
temperature uniquely defined for all particle species, but rather a statistical
average, may be considered to be appropriate for many purposes. However,
in the search for evidence for a critical point such a 
difference has substantial consequences for expected properties of
net charge fluctuations as the size of the critical region, in which charge
fluctuations may become large, may well be only of that order 
\cite{Schaefer:2006ds}. A 10~MeV difference between $T_c$ and $T_f$ thus may 
decide whether or not freeze-out happens in the critical region.

\begin{figure}[t]
\begin{center}
\hspace*{-0.1cm}\includegraphics[width=72mm]{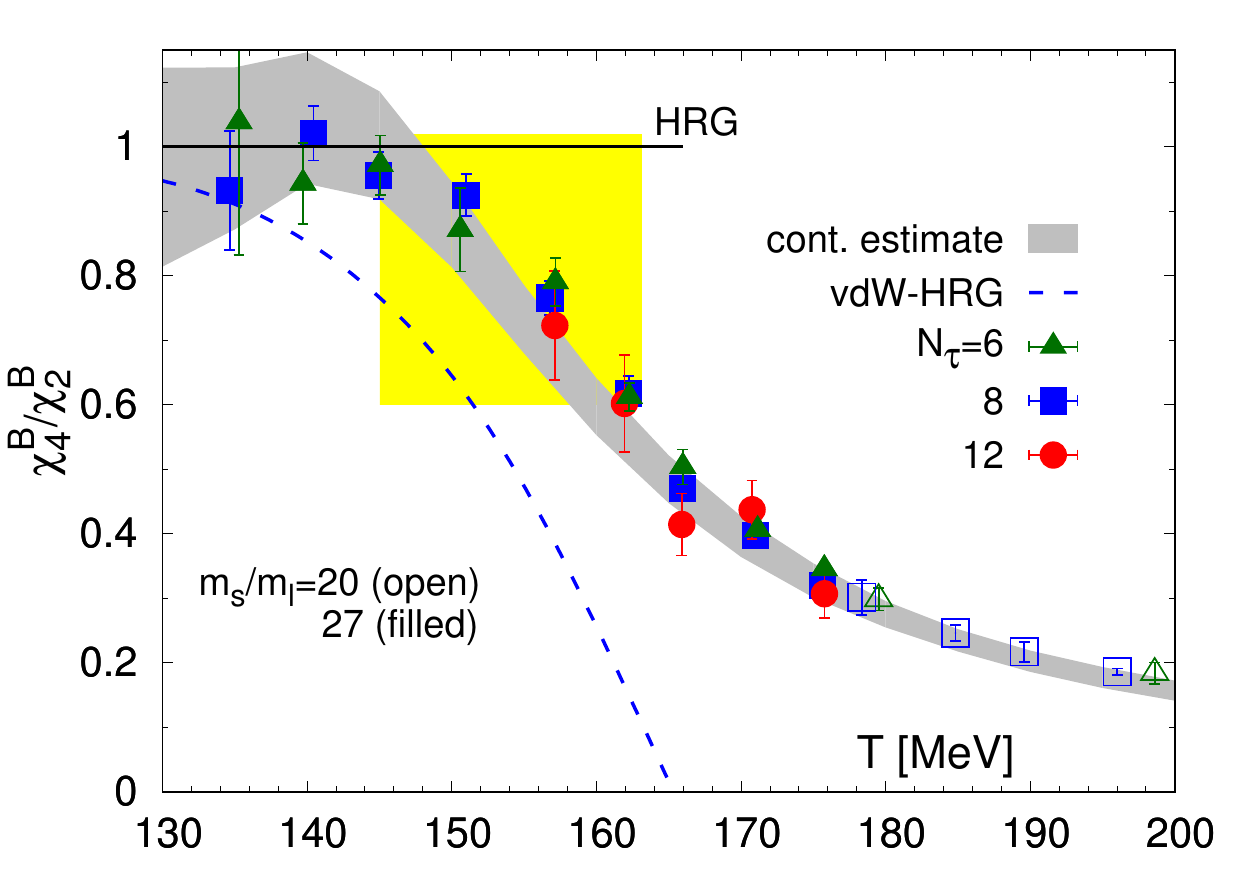}\hspace*{-0.1cm}
\includegraphics[width=72mm]{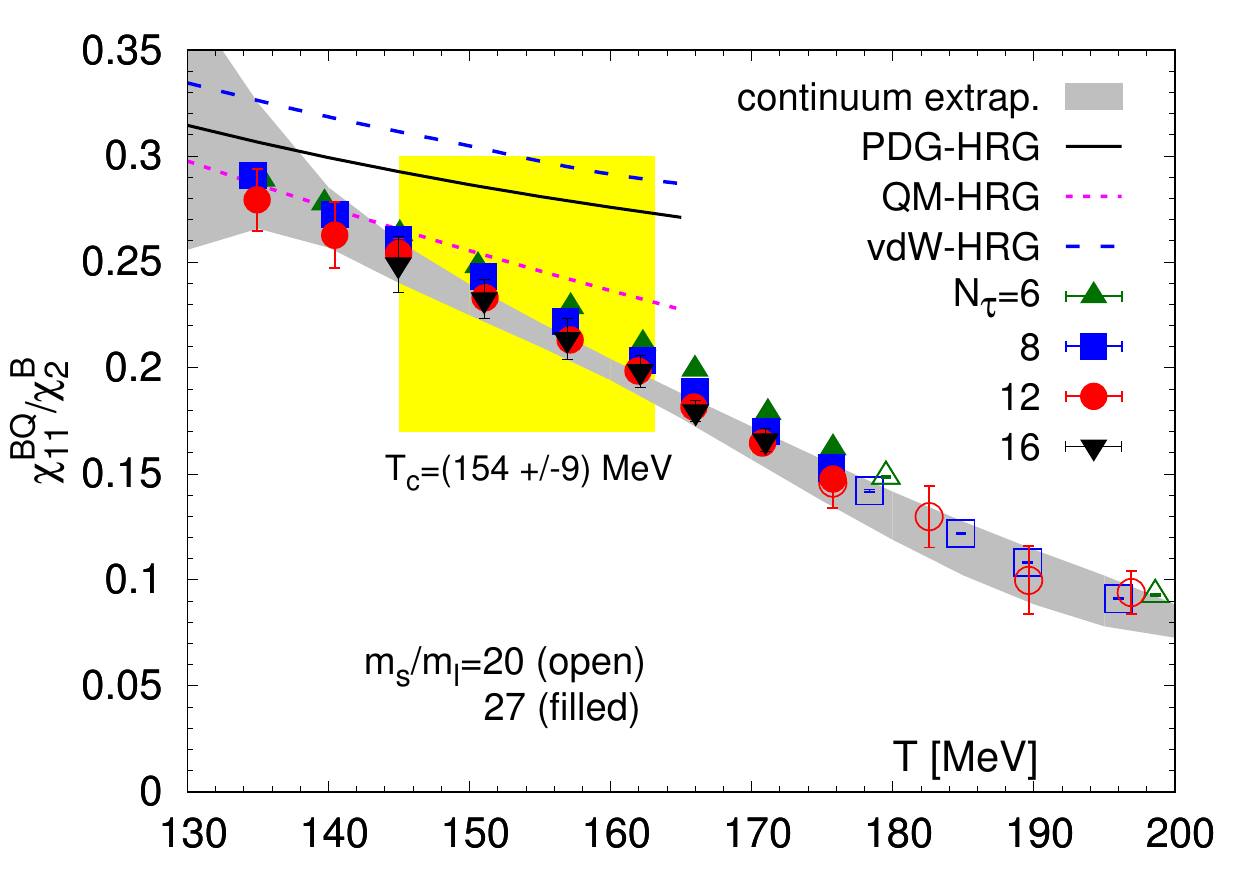}
\caption{
The ratio of fourth and second order cumulants of net baryon-number 
fluctuations (left) and the correlations between
net electric charge and net baryon-number fluctuations
normalized to the second order cumulant of net baryon-number fluctuations
(right). The lines show results obtained in various hadron resonance gas
model calculations (see text).
\vspace*{-0.5cm}
}
\label{fig:chiratios}
\end{center}
\end{figure}

Cumulants of net charge fluctuations and correlations among
fluctuations of different conserved charges, i.e. baryon number ($B$),
electric charge ($Q$) and strangeness ($S$) can be obtained as 
derivatives of the logarithm of the QCD partition function \cite{Bazavov:2017dus},
\begin{equation}
\chi_{n}^{X} = \left. \frac{1}{VT^3} 
\frac{\partial^{n} \ln Z(V,T,\vec{\mu})}{\partial X^n}
\right|_{\vec{\mu}=0}\;\; ,\;\; 
\chi_{nm}^{XY} = \left. \frac{1}{VT^3} 
\frac{\partial^{(n+m)} \ln Z(V,T,\vec{\mu})}{\partial X^n\  \partial Y^m}
\right|_{\vec{\mu}=0}\;\; ,\;\; X,\ Y=B,\ Q,\ S \;\; ,
\label{cumulants}
\end{equation}
with $\vec{\mu} \equiv (\mu_B,\ \mu_Q,\ \mu_S)$ denoting the three
chemical potentials connected with the conserved charges.
As can be seen in Fig.~\ref{fig:chiratios}  ratios of cumulants of 
net charge fluctuations change rapidly in the crossover region.
The ratio of $4^{th}$ and $2^{nd}$ order net baryon-number fluctuations,
$\chi_4^B/\chi_2^B$,
changes from almost unity at $T=145$~MeV to about 0.5 at $T=165$~MeV
(see Fig.~\ref{fig:chiratios}~(left)). In the temperature range identified
by ALICE as the freeze-out region, $T_f=156(2)$~MeV, the ratio is 
$\chi_4^B/\chi_2^B\simeq 0.75$. The situation is similar for ratios of 
conserved charge correlations. For the ratio of $4^{th}$ and $2^{nd}$ order 
cumulants characterizing correlations between
net strangeness and net baryon-number fluctuations one finds
$\chi_{31}^{BS}/\chi_{11}^{BS}\simeq 0.74$ at $T=155$~MeV, while
this ratio drops to $\sim 0.36$ at $T=165$~MeV. This is in contrast to
HRG model calculations with point-like non-interacting hadrons, where
these ratios are unity irrespective of the particle content in the hadron
spectrum.

Also on the level of second order cumulants, which currently get
measured by the ALICE Collaboration \cite{Rustamov:2017lio}, 
differences between HRG model 
calculations and QCD results are already significant. In 
Fig.~\ref{fig:chiratios}~(right) we show the ratio, $\chi_{11}^{BQ}/\chi_2^B$,
which in resonance gas models for non-interacting, point-like hadrons has the
interpretation of the relative contribution of charged baryons to the 
total baryon contribution of the pressure. As can be seen at $T=165$~MeV
the ratio $\chi_{11}^{BQ}/\chi_2^B$ is about 50\% smaller than predicted
by HRG model calculations based on the experimentally known hadron spectrum
(PDG-HRG). Model calculations that include additional strange baryon
resonances predicted in QCD motivated quark model calculations (QM-HRG)
provide a better approximation to QCD calculations but still over-predict
the ratio $\chi_{11}^{BQ}/\chi_2^B$ for $T\gsim 150$~MeV. A similar
behavior has been found for the correlations between net baryon-number
and net strangeness fluctuations, $\chi_{11}^{BS}$, normalized to either 
$\chi_2^B$ or $\chi_2^S$. This may be taken as evidence for the importance
of additional strange baryon resonance contributing significantly
to the thermodynamics close to the chiral transition. 

There is, of course,
also the possibility that additional interactions among hadrons need to
be taken into account in order to improve agreement between QCD and hadron
gas calculations at low temperature. However, as is also evident from
Fig.~\ref{fig:chiratios}, it seems that neither
taking into account 
repulsive interactions through
the finite volume of baryons \cite{Andronic:2012ut} 
nor an additional
attractive contribution through
a van der Waals interaction (vdW-HRG) can improve the 
validity range of HRG model based calculations \cite{Vovchenko:2016rkn}. 

\section{Cumulants of conserved charge fluctuations at non-zero net
baryon number} 

\begin{figure}[t]
\begin{center}
\hspace*{-0.1cm}\includegraphics[width=72mm]{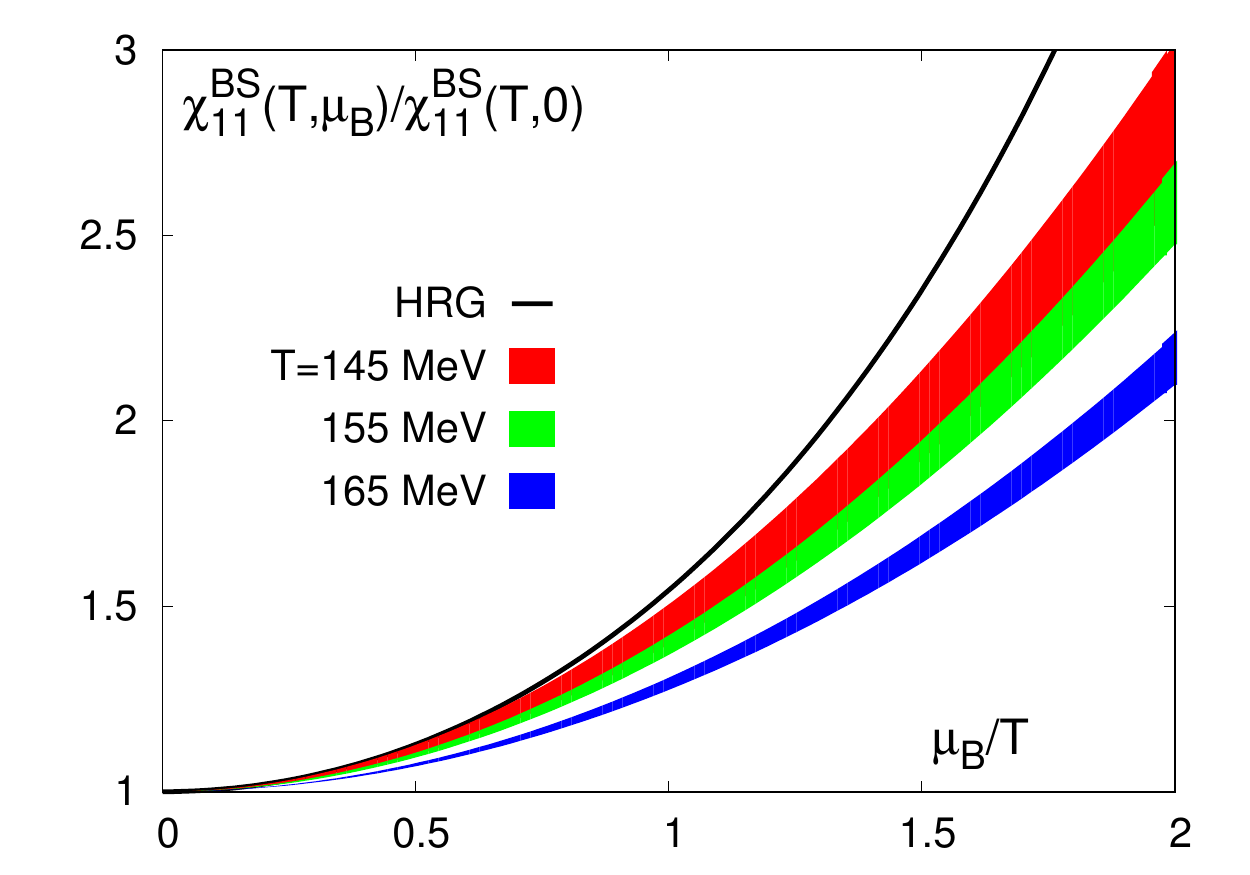}\hspace*{-0.1cm}
\includegraphics[width=72mm]{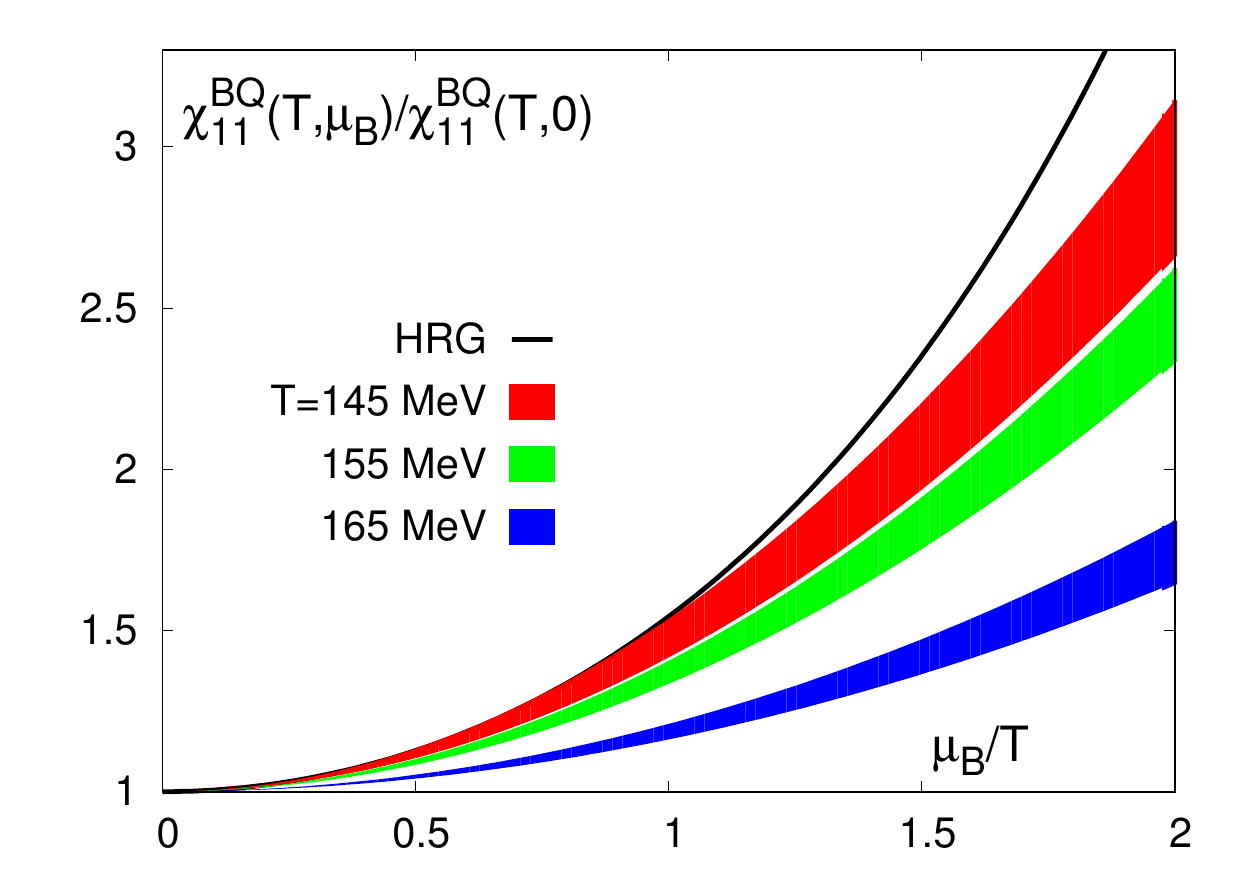}
\caption{ Correlation between net baryon-number fluctuations
and net strangeness fluctuations (left) as well as net electric charge
fluctuations (right) for three values of the temperature. Shown are results
from a Taylor expansion up to ${\cal O}(\mu_B^2)$ with $\mu_Q=\mu_S=0$.
Also shown are results for a hadron resonance gas in the Boltzmann 
approximation.
\vspace*{-0.5cm}
}
\label{fig:chi42}
\end{center}
\end{figure}

Ratios of $4^{th}$ and $2^{nd}$ order cumulants such as 
$\chi_4^B/\chi_2^B$ (Fig.~\ref{fig:chiratios}~(left)) 
or $\chi_{31}^{BX}/\chi_{11}^{BX}$, $X=Q,\ S$,
also give the leading ${\cal O}(\mu_B^2)$
corrections to the $2^{nd}$ order cumulants themselves. For 
$\mu_Q=\mu_S=0$ one has 
\begin{equation}
\chi_{2}^{B}(T,\mu_B) = \chi_{2}^{B} + \frac{1}{2}\chi_{4}^{B} 
\left( \frac{\mu_B}{T} \right)^2+ {\cal O}(\mu_B^4)\;\;\;\; ,\;\;\;\;
\chi_{11}^{BX}(T,\mu_B) = \chi_{11}^{BX} + \frac{1}{2}\chi_{31}^{BX} 
\left( \frac{\mu_B}{T} \right)^2+ {\cal O}(\mu_B^4)  \;\;,\;\; X=Q, \; S \; .
\label{expansion}
\end{equation}
As these ratios are smaller than the corresponding HRG
values, this also means that differences between lattice QCD results for
these cumulants and HRG model calculations increase with $\mu_B$. Some 
results for $\chi_{11}^{BS}$ and $\chi_{11}^{BQ}$ at three values
of the temperature are shown in Fig.~\ref{fig:chi42}. Obviously 
deviations from HRG model calculations become large for $\mu_B/T \gsim 1.5$,
which in the BES at RHIC corresponds to $\sqrt{s_{NN}} \lsim 15$~GeV.

The situation is similar for ratios of cumulants
of net baryon-number fluctuations. 
For instance,
\begin{equation}
\frac{\chi_4^B(T,\mu_B)}{\chi_2^B(T,\mu_B)}
= \frac{\chi_4^B}{\chi_2^B}\left( 1 + \frac{1}{2} \left( 
\frac{\chi_6^B}{\chi_4^B} - \frac{\chi_4^B}{\chi_2^B} \right)
\left( \frac{\mu_B}{T} \right)^2 +  {\cal O}(\mu_B^4) \right)
\;\;\; {\rm for} \;\;\; \mu_Q=\mu_S=0\; .
\label{B42}
\end{equation}
is unity in HRG model calculations with non-interacting, point-like hadrons
for all values of $\mu_B$. In QCD, however, the ${\cal O}(\mu_B^2)$
expansion coefficient in Eq.~\ref{B42} is negative
for $T\gsim 150$~MeV \cite{Karsch:2015nqx}. This indicates that
the ratio $\chi_4^B(T,\mu_B) / \chi_2^B(T,\mu_B)$ becomes smaller
with increasing $\mu_B$. Even if the ratio is close to HRG model
results at $\mu_B=0$, it thus will further deviate from this with
increasing $\mu_B$. This is consistent with the trend found by 
STAR for the corresponding kurtosis ratio of net proton-number 
fluctuations, $\kappa_P\sigma^2_P=\chi_4^P/\chi_2^P$ \cite{Thader:2016gpa}.

\section{Conclusions}
Many ratios of $4^{th}$ and $2^{nd}$ order cumulants of conserved charge
fluctuations calculated in QCD
agree with HRG model calculations within 25\% or better at temperatures
below $T=155$~MeV. For observables including strangeness fluctuations this
often requires to take into account contributions from additional strange
hadrons not listed in the Particle Data Tables.  At $T=165$~MeV  
HRG model calculations often deviate from QCD results by more than 50\%
even for observables that only involve quadratic charge fluctuations.

\vspace*{0.2cm}
\noindent
{\it Acknowledgements:}
This work was supported through Contracts No. DE-SC001270 with the
U.S. Department of Energy and  No. 05P15PBCAA with the 
German Bundesministerium f\"ur Bildung und Forschung. 
\bibliographystyle{elsarticle-num}
\bibliography{<your-bib-database>}

\begin{thebibliography}{00}
\bibitem{Ding:2015ona} 
  H.~T.~Ding, F.~Karsch and S.~Mukherjee,
  Int.\ J.\ Mod.\ Phys.\ E {\bf 24}, no. 10, 1530007 (2015)
  [arXiv:1504.05274 [hep-lat]].

\bibitem{Gunther:2016vcp} 
  J.~Gunther {\it et al.}, 
  EPJ Web Conf.\  {\bf 137}, 07008 (2017)
  [arXiv:1607.02493 [hep-lat]].

\bibitem{Bazavov:2017dus}
  A.~Bazavov {\it et al.},
  Phys.\ Rev.\ D {\bf 95} 054504 (2017),
  [arXiv:1701.04325 [hep-lat]].

\bibitem{Bazavov:2011nk} 
  A.~Bazavov {\it et al.},
  Phys.\ Rev.\ D {\bf 85}, 054503 (2012)
  [arXiv:1111.1710 [hep-lat]].

\bibitem{Bazavov:2014pvz}
  A.~Bazavov {\it et al.}  [HotQCD Collaboration],
  Phys.\ Rev.\ D {\bf 90}, 094503 (2014)
  [arXiv:1407.6387 [hep-lat]].

\bibitem{Floris:2014pta}
  M.~Floris,
  Nucl.\ Phys.\ A {\bf 931}, 103 (2014)
  [arXiv:1408.6403 [nucl-ex]].
\bibitem{Das:2014qca}
  S.~Das [STAR Collaboration],
  EPJ Web Conf.\  {\bf 90}, 08007 (2015)
  [arXiv:1412.0499 [nucl-ex]].
\bibitem{Becattini:2016xct}
  F.~Becattini, J.~Steinheimer, R.~Stock and M.~Bleicher,
  Phys.\ Lett.\ B {\bf 764}, 241 (2017)
  [arXiv:1605.09694 [nucl-th]].

\bibitem{Schaefer:2006ds}
  B.~J.~Schaefer and J.~Wambach,
  Phys.\ Rev.\ D {\bf 75}, 085015 (2007)
  [hep-ph/0603256].

\bibitem{Rustamov:2017lio} 
  A.~Rustamov [ALICE Collaboration],
  arXiv:1704.05329 [nucl-ex], these proceedings.

\bibitem{Andronic:2012ut} 
  A.~Andronic, P.~Braun-Munzinger, J.~Stachel and M.~Winn,
  Phys.\ Lett.\ B {\bf 718}, 80 (2012)
  [arXiv:1201.0693 [nucl-th]].
\bibitem{Vovchenko:2016rkn} 
  V.~Vovchenko, M.~I.~Gorenstein and H.~Stoecker,
  Phys.\ Rev.\ Lett.\  {\bf 118}, 182301 (2017)
  [arXiv:1609.03975 [hep-ph]].

\bibitem{Karsch:2015nqx} 
  F.~Karsch {\it et al.},
  Nucl.\ Phys.\ A {\bf 956}, 352 (2016)
  [arXiv:1512.06987 [hep-lat]].

\bibitem{Thader:2016gpa}
J.~Th\"ader [STAR Collaboration],
  Nucl.\ Phys.\ A {\bf 956}, 320 (2016)
\end{thebibliography}

\end{document}